\documentclass[aps,prl,twocolumn,showpacs,superscriptaddress,amsmath,amssymb]{revtex4}
\usepackage{epic}
\usepackage{curves}
\usepackage{graphicx}
\begin{document}

%\preprint{}

\title{Low Energy States of $^{81}_{31}$Ga$_{50}$ :\\
Elements on the Doubly-Magic Nature of $^{78}$Ni}


\author{D. Verney}
\email[]{verney@ipno.in2p3.fr}
%\thanks{}
\affiliation{Institut de Physique Nucl\'eaire CNRS-IN2P3/Univ. Paris Sud-XI, F-91406 Orsay Cedex, France}
\affiliation{GANIL, BP 55027, F-14076 Caen Cedex 5, France}
\author{F. Ibrahim}
\author{C. Bourgeois}
\author{C. Donzaud}
\author{S. Essabaa}
\affiliation{Institut de Physique Nucl\'eaire CNRS-IN2P3/Univ. Paris Sud-XI, F-91406 Orsay Cedex, France}
\author{S. Gal\`es}
\author{L. Gaudefroy}
\affiliation{GANIL, BP 55027, F-14076 Caen Cedex 5, France}
\affiliation{Institut de Physique Nucl\'eaire CNRS-IN2P3/Univ. Paris Sud-XI, F-91406 Orsay Cedex, France}
\author{D. Guillemaud-Mueller}
\author{F. Hammache}
\author{C. Lau}
\author{F. Le Blanc}
\author{A.C. Mueller}
\author{O. Perru}
\author{F. Pougheon}
\author{B. Roussi\`ere}
\author{J. Sauvage}
\affiliation{Institut de Physique Nucl\'eaire CNRS-IN2P3/Univ. Paris Sud-XI, F-91406 Orsay Cedex, France}
\author{O. Sorlin}
\affiliation{GANIL, BP 55027, F-14076 Caen Cedex 5, France}
\affiliation{Institut de Physique Nucl\'eaire CNRS-IN2P3/Univ. Paris Sud-XI, F-91406 Orsay Cedex, France}
\collaboration{the PARRNe Collaboration}
%\thanks{\textbf{A}cc\' el\' erateur \textbf{L}in\' eaire aupr\` es du \textbf{T}andem %d'\textbf{O}rsay~: former PARRNe Collaboration}
%\homepage{http://ipnweb.in2p3.fr/tandem-alto/}
\affiliation{Institut de Physique Nucl\'eaire CNRS-IN2P3/Univ. Paris Sud-XI, F-91406 Orsay Cedex, France}

\date{\today}

\begin{abstract}
Excited levels were attributed to $^{81}_{31}$Ga$_{50}$ for the first time which were fed in the $\beta$-decay of its mother nucleus $^{81}$Zn produced in the fission of $^{nat}$U using the ISOL technique. We show that the structure of this nucleus is consistent with that of the less exotic proton-deficient $N=50$ isotones within the assumption of strong proton $Z=28$ and neutron $N=50$ effective shell effects.
\end{abstract}

%\keywords{}

\maketitle

%\section{}
%\subsection{}
%\subsubsection{}
The persistence of the magic character of the number of nucleons of one family, protons or neutrons, while the population of the other is varied has been clearly traced back to the interplay between the monopole part of the effective nuclear interaction and correlations of other nature.  The monopole part is responsible for the change in the effective single particle energies (SPE) and it has been shown that its tensor term is to play a significant role \cite{otsuka}. The main origin of the competing correlations of other nature are pairing correlations and quadrupole coherence \cite{zukerT}. This interplay can enhance the magic character of a number which does not belong to the historical sequence 2, 8, 20, 28, 50, 82, 126 like $N=40$ in the case of $^{68}$Ni \cite{sorlin}. On the opposite, it can lead to complete inversion of the 0p-0h and 2p-2h configurations leading to the vanishing of its magic character like for $N=20$ in the case of $^{32}$Mg \cite{DGM}. Those effects have been seen however to be very localized within a small number of nucleons and it is known that the robustness of spin-orbit (SO) closures prevents them from apparent eradication while closures of other origins are more fragile \cite{zuker}. 
It means for instance that no 2p-2h configuration has been observed as main component of the ground state (g.s.) in nuclei with nucleon numbers of the SO sequence 28, 50, 82, 126. For how long this will stand remains an open question~: experimental evidence found in the most exotic part of the nuclide chart tends indeed to accumulate showing that such configurations come very close to the g.s. like in some $N=28$ isotones \cite{sohler,grevy} or in some $Z=82$ (lead) isotopes \cite{plomb}. 
%Besides, it has been shown that evolution of the nuclear mean field itself toward the drip lines %can give rise to new (non historical) genuine magic numbers like $N=16$ \cite{ozawa}. 
From all these considerations the problem of $^{78}$Ni is easy to formulate~: with two SO historical magic numbers $Z=28$ and $N=50$ it should be a doubly magic nucleus but a special one since, being far remotely situated from stability, the way the monopole-quadrupole interplay will drive its structure remains an open question (and a challenging one). The study of the structure of $^{78}$Ni itself belongs to the medium-term development of the next generation of radioactive ion beam facilities. Nevertheless the study of the two magic numbers $Z=28$ and $N=50$ in its vicinity has already been undertaken for a long time and is still the object of active experimental and theoretical research. The $Z=28$ (nickel) isotopes are well known for being magic nuclei, however the experimentally observed \cite{franchoo} evolution of the effective SPE of the proton $1f_{5/2}$ in Cu isotopes could be considered as a preliminary indication of the reduction of the $Z=28$ shell gap, depending on the simultaneous evolution of the effective SPE of the $\pi 1f_{7/2}$ orbital. How the $N=50$ shell gap evolves below $Z=38$ was somewhat debated  \cite{zhang,prevost} but it has been shown recently that the associated shell effect should dominate the structure of the low lying states in nuclei with $Z$ as low as 32 (Ge) \cite{perru}. New experimental data on proton-deficient $N=50$ isotones is clearly needed in order to clarify the situation.
In this Letter we report on the first discovery of the low-lying structure of $^{81}$Ga which, with $Z=31$ and $N=50$, has only three protons more than $^{78}$Ni.\\
In our experiment, $\gamma$-rays de-exciting levels in $^{81}$Ga fed in the $\beta$-decay of $^{81}$Zn were observed. 
The sources of $^{81}$Zn ($T_\frac{1}{2}=290\pm 50$ ms) were obtained at the PARRNe mass-separator operating on-line at the 15MV MP-Tandem of the Institute of Nuclear Physics, Orsay. A $^{nat}$U target of approximative mass 75 g made of a series of UC$_x$ pellets heated at $\simeq 2000^{\circ}\mbox{C}$ was associated with a hot ($\simeq 1800^{\circ}\mbox{C}$) plasma ion-source of the ISOLDE MK5 type \cite{sundell} and exposed to the neutron flux generated by the reaction of the 26 MeV deuteron beam delivered by the Tandem hitting the target container.
The ions, extracted at 30 kV from the source and magnetically mass separated were deposited on a Al-coated Mylar tape close to the detection system. 
The rate of implanted $^{81}$Zn was estimated to a few tens per second. 
The $\gamma$-detection system consisted in two coaxial large volume HPGe detectors of the EUROGAM phase I type (70\% relative efficiency) issued from the French/U.K. (IN{\small 2}P{\small 3}/EPSRC) Loan Pool. They were placed in $180^{\circ}$ geometry close to the point at which the beam was deposited onto the tape (collection point). The energy resolution achieved with these detectors was of the order of 2.3 keV at 1 MeV. The collection point was surrounded by a tube-shaped plastic scintillator for $\beta$-detection. Enhancement of the activity of interest as respect to the longer lived activities from the other collected isobars and the consequent Compton background was made by moving cyclically (every 2100 ms) the tape after a short build-up (900 ms) and decay time (1200 ms). The $Z$-identification of the $\gamma$-rays was provided by the analysis of the evolution of their activities during the decay part of the cycle. More details on the experimental set-up and procedure can be found in \cite{perru} and Refs. therein.
\begin{figure}
\resizebox{0.45\textwidth}{!}{
\includegraphics*{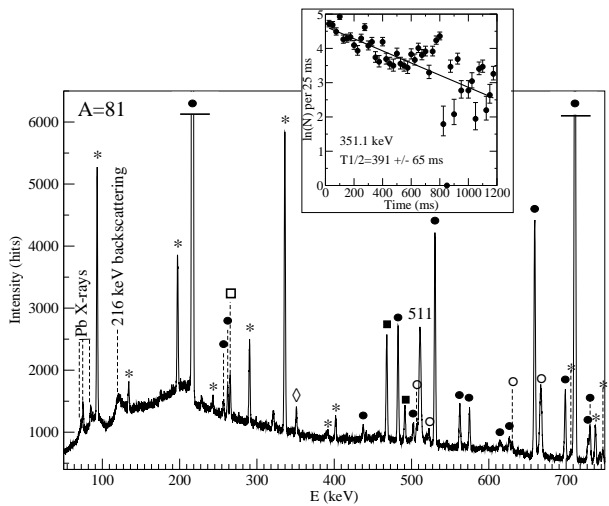}
}
\caption{\label{fig:spectre} Part of the $\beta$-gated $\gamma$-spectrum recorded at mass 81. The observed $\gamma$-lines have been identified as transitions fed in the $\beta$-decays of $^{81}$Ga(\textbullet -symbol), $^{81}$Ge ($\ast$-symbol), $^{81}$As ($\blacksquare$-symbol) and the $\beta$-n decay of $^{81}$Ga ($\square$-symbol, $P_n=11.9\%$). The activity of $^{132}$I ($\circ$-symbol) is also identified, it comes from a previous setting on mass 132 which is oftenly used as a reference in our experiments. The new line at 351.1 keV ($\diamondsuit$-symbol) is clearly visible and well isolated. The projection on the time scale of the background substracted events in this peak is shown in the insert.}
\end{figure}
Part of the $\gamma$-spectrum recorded at mass 81 is displayed in Fig. \ref{fig:spectre} along with the identification of the $\gamma$-lines. It is seen that the activity is dominated by that of $^{81}$Ga. In spite of this, a $\gamma$-line at 351.1 keV previously not reported at $A=81$ is clearly visible. The fit of the evolution in time of the $\gamma$-intensity during the decay part of the counting cycle (see the insert in Fig. \ref{fig:spectre}) gives a half-life value of 391(65)ms which is consistent with the known value for $^{81}$Zn T$_{1/2}=290(50)$ ms \cite{NDS}~: it was then attributed to a transition in $^{81}$Ga.
\begin{figure}
\resizebox{0.45\textwidth}{!}{
\includegraphics{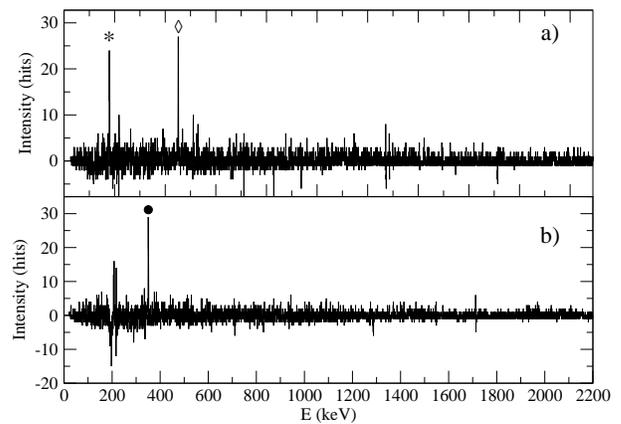}
}
\caption{\label{fig:coinc} Part a)~: $\gamma$-ray spectrum in coincidence with the peak at 351.1 keV. A narrow line is clearly observable at 451.7 keV ($\diamondsuit$-symbol) which was barely visible in the direct spectra. The wider line around 179 keV ($\ast$-symbol) is due to the $180^\circ$-backscatter of the 530.22 keV $\gamma$-ray from the $^{81}$Ga decay. Part b)~: $\gamma$-ray spectrum in coincidence with the peak at 451.7 keV. The 351.1 keV peak is clearly visible (\textbullet -symbol), the negative-positive oscillation around 200 keV has a similar origin as $\ast$ in part a).}
\end{figure}
\begin{figure}
\resizebox{0.22\textwidth}{!}{
\includegraphics{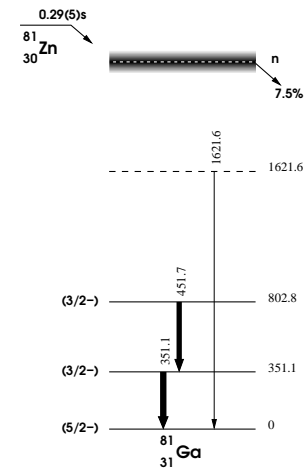}
}
\caption{\label{fig:level} Tentative experimental level scheme for $^{81}$Zn. The order of the two transitions seen in coincidence is based on intensity consideration. The proposed spin assignations come from the discussion in the present Letter. The 1621.6 keV level is proposed after the observation of a very weak peak in the spectra, not discussed in the text.}
\end{figure}
The statistics was sufficient to allow coincidence observation~: as can be seen in Fig. \ref{fig:coinc} a 451.7 keV $\gamma$-line was observed in coincidence with the 351.1 keV line which establishes the existence of a 802.8 keV level in the scheme of $^{81}$Zn (see Fig. \ref{fig:level}).\\
In a first attempt to understand the nature of the observed states we compared our results to shell-model calculations. The calculations were performed using the ANTOINE code from the Strasbourg group \cite{antoine}. The model space chosen for the calculations consists in an inert $^{78}$Ni core and the proton single particle states from $Z=28$ to $Z=50$ \textit{i.e.} $\left\{1f_{5/2},2p_{3/2},2p_{1/2},1g_{9/2}\right\}$. From the introductory remarks this should be the natural valence space where the low-lying structure of $^{81}$Ga develops and it is actually the one which has been traditionally (and successfully) used in shell-model descriptions of the $N=50$ isotones. We used two different sets of SPE and two-body matrix elements (TBME) for the effective interaction, both were determined as free parameters in a least square fit procedure from experimental levels and binding energies~:
%the set denoted as \textquotedblleft JW\textquotedblright\ in the following, from Ref. \cite{jw} %and the set \textquotedblleft jj4pna\textquotedblright\ from Ref. \cite{liset}.
the one proposed by Ji and Wildenthal (JW) some time ago \cite{jw} and the one, which includes new experimental levels accumulated since that time, proposed more recently by Lisetskiy \textit{et al.} (jj4pna) \cite{liset}.
\begin{figure}
\resizebox{0.45\textwidth}{!}{
\includegraphics*{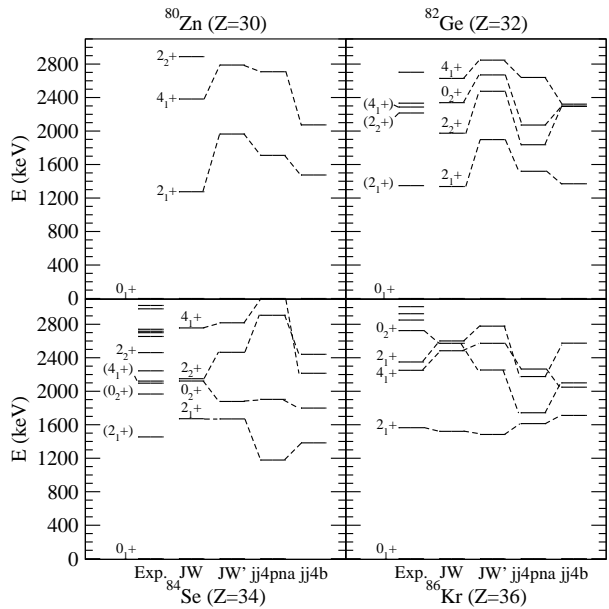}
}
\caption{\label{fig:pairs} Experimental and calculated spectra for the last stable ($^{86}$Kr) and proton-deficient even $N=50$ isotones. \textquotedblleft Exp.\textquotedblright\ stands for experimental levels, \textquotedblleft JW\textquotedblright\ for results of shell-model calculations using the interaction from \cite{jw}, \textquotedblleft JW$^\prime$\textquotedblright\ the same with modified pairing (see text), \textquotedblleft jj4pna\textquotedblright\ and \textquotedblleft jj4b\textquotedblright\ those from \cite{liset} and \cite{lb2} respectively.}
\end{figure}
\begin{figure}
\resizebox{0.45\textwidth}{!}{
\includegraphics*{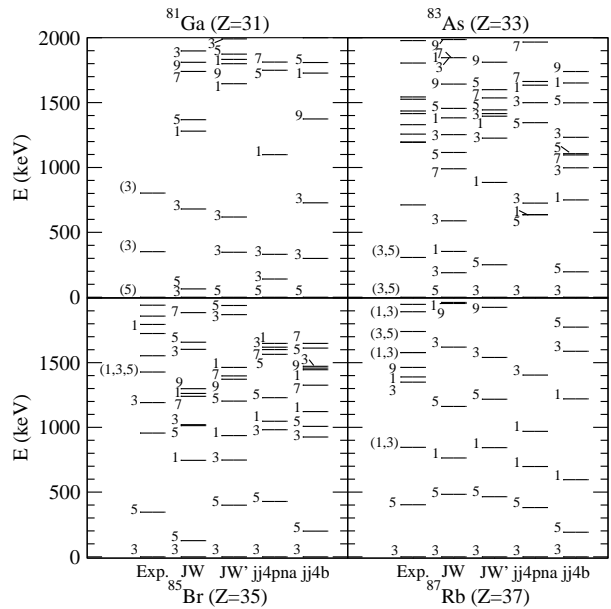}
}
\caption{\label{fig:impairs} Experimental and calculated spectra for the last stable ($^{87}$Rb) and proton-deficient odd $N=50$ isotones, spin values are multiplied by two, only the three first J$^\pi=3/2^-, 5/2^-, 7/2^-$ and $9/2^-$ calculated levels are displayed. Notation for the interactions is the same as in Fig. \ref{fig:pairs}.}
\end{figure}
As can be seen from Fig. \ref{fig:pairs} a good agreement is obtained between theory and experiment for the energy of the $2^+_1$ excited states with both interactions while for the rest of the level scheme the results tend to differ strongly. In particular the agreement is excellent \textemdash\ especially for the JW interaction \textemdash\ for the surprisingly low energy of the $2^+_1$ state of the last known even proton-deficient isotone $^{82}_{32}$Ge ($^{78}$Ni$+4$ protons). However when considering the results for the odd-nuclei (Fig. \ref{fig:impairs}) the agreement deteriorates rapidly when going away from stability. Difficulties already show up in $^{83}_{33}$As ($^{78}$Ni$+5$ protons) where the experimentally well established 711-keV level could not be accounted for nor with the JW interaction \textemdash\ this was considered as a major issue at the time $^{83}_{33}$As was studied for the first time \cite{winger} \textemdash\  nor with the more recent jj4pna interaction which predicts a compact group of three states around 700 keV and no intermediate state around 300 keV. Besides, the two interactions does not agree on the nature of the $^{83}_{33}$As g.s.~: it is found $J^\pi=5/2^-$ with JW and $3/2^-$ with jj4pna. Since the quality of the agreement between theory and experiment is already doubtful for $Z=33$ it is even more difficult to raise a conclusion for $Z=31$.
This can be understood as due to the fact that the TBME  involving the $1f_{5/2}$ proton states which determine the details of the structure of the most proton-deficient isotones are necessarily the less safely determined parameters.
It is easy to see for instance that the term $\left\langle 1f_{5/2} 1f_{5/2} \right|V_{12}\left| 1f_{5/2} 1f_{5/2}\right\rangle_{J=0\ T=1}$ of the JW interaction (Tab. 1 in Ref. \cite{jw}) which corresponds to the pairing energy between $1f_{5/2}$ protons is not well converged in the fitting process and has a strangely low value~: $-0.7854$ MeV against $-1.5115$ MeV in \cite{liset}. It is then tempting to modify the TBME involving the $\pi 1f_{5/2}$ orbit following empirical prescriptions in the light of our new experimental data. For this we make the voluntarily oversimplifying hypothesis (in order to keep a clear physical image) that the main configuration in the g.s. of the nuclei with $29\le Z \le 34$ is $\pi 1f_{5/2}^{n}$ ($1\le n \le 6$). The configuration energy for $n$ identical particles with $j= 5/2$ coupled to the total angular momentum $J$ and seniority $v$ is a closed formula valid for \textit{any} two body residual interaction \cite{talmiunna}~:
%(seniority being necessarily conserved within a configuration with $j\le 7/2$).
\begin{multline*}
\label{equation1}
%\begin{split}
\left\langle j^nJv\right|\sum^{n}_{i<k}V_{ik}\left|j^nJv \right\rangle=n\varepsilon_{f_{5/2}}+\frac{1}{2}n(n-1)a\\
+\left(J(J+1)-\frac{35}{4}n\right)b
+\frac{1}{2}(n-v)(8-n-v)c
%\end{split}
\end{multline*}
with four parameters $a,b,c$ and $\varepsilon_{f_{5/2}}$, the last one being
the SPE of the proton $1f_{5/2}$ in the core mean field. As a first guess, we assumed that the g.s. of $^{79}$Cu, $^{81}$Ga and $^{83}$As were the $J_v=5/2_1$ members of the $1f_{5/2}^{n}$ configurations, which, as we shall see later, is not totally exact. We determined three relations between the parameters by fitting the quadratic dependence in $n$ of the binding energies (BE) of these nuclei taken relative to $^{78}$Ni. The BE values were taken from \cite{audi}, in the case they were not measured the estimated values were used, which is far sufficient in the framework of the crude hypothesis of pure configuration states. 
The fourth relation was provided by the energy separation between the ground and the second excited state of $^{81}$Ga taken from the present work (802.8 keV)~: the second excited state in $^{81}$Ga was assumed to be the $J^\pi_v=3/2^-_3$ member of the $\pi 1f_{5/2}^{3}$ configuration (the last member $J_v=9/2_3$ should be weakly fed in the $\beta$-decay of $^{81}$Zn which has most probably a $5/2^+$ g.s. of neutron $2d_{5/2}$ nature).
By simply resolving the system, the three diagonal $1f_{5/2}$ TBME plus the BE of the $\pi 1f_{5/2}$ single particle were determined uniquely. The value obtained for $\left\langle \frac{5}{2}\frac{5}{2}\right|V_{12}\left|\frac{5}{2}\frac{5}{2} \right\rangle_{J=0\ T=1}$ is $-1.478$ MeV~: it is about twice the JW value and rather close to the jj4pna value. In order to reconnect properly the $\pi 1f_{5/2}$ orbit to the rest of the valence space it was also necessary to slightly modify the TBME determining the position of the $\pi 1f_{5/2}$-centro\"\i de and those which govern the diffusion of pairs from the $\pi 1f_{5/2}$ orbit to the close lying $\pi 2p_{3/2}$. This gives a total of 8 TBME, all related to pairing properties, which have been modified as respect to the original values of JW. The resulting spectra are shown in Figs. \ref{fig:pairs} and \ref{fig:impairs} (JW$^\prime$). As can be seen in Fig. \ref{fig:impairs}, a clear improvement is obtained for the lower part of the spectra of $^{81}$Ga and $^{83}$As. The major result is that, for $^{83}$As the good number of levels below the 1.5 MeV group is obtained for the first time. In particular the intermediate level which is observed at 711 keV in this nucleus would correspond to the $1/2^-$ state calculated at 884 keV. Besides, the $^{83}$As g.s. becomes $3/2^-$ with a $2p_{3/2}$ q.p. nature and the $5/2^-_{v=1}$ state of the $1f_{5/2}^{-1}$ configuration becomes the first excited state. In $^{81}$Ga the $5/2^-_{v=1}$ member of the $1f_{5/2}^{3}$ configuration becomes the g.s. while
%, and consistently with %the experimental input, the second excited state is mainly the %$3/2^-_{v=2}$ member of the %$1f_{5/2}^{3}$ configuration. This corresponds to a permutations of %these two states as respect to the JW calculation which predicted an unnatural $3/2^-_{v=2}$ %g.s. From the JW$^\prime$ interaction, 
the $2p_{3/2}$ q.p. becomes the first excited state. This provides a natural and simple explanation for a possible change of the g.s. spin value between $^{83}$As and $^{81}$Ga as being due to the lowering of the proton Fermi level from the $2p_{3/2}$ to the $1f_{5/2}$ orbitals while the JW calculation predicted an unnatural $3/2^-_{v=2}$ g.s for $^{81}$Ga. Our trivial manipulation of the TBME tends to rigidify the structure of the even-nuclei (see fig. \ref{fig:pairs}) while it improves the results on odd-nuclei. This could be understood as the evidence that part of the correlations which contributes to lower the $2^+_1$ energies toward $^{78}$Ni cannot be reproduced in such a limited valence space and may come typically from the diffusion of pairs across the $Z=28$ shell-gap. However a proper modification of the whole set of the interaction parameters has been performed recently by A. F. Lisetskiy \textit{et al.} (unpublished \cite{lb2})
including, in particular, data on $^{84}$Se in the fit along with about 400 other binding energy
and energy level data.
The resulting spectra are displayed in Figs. \ref{fig:pairs} and \ref{fig:impairs} (jj4b) where it is seen that the agreement with the experimental spectra of the even-nuclei is very good and the description of the odd-nuclei is improved in a similar way as with the JW$^\prime$ set.
%As could be expected from a simple increase of the pairing strength in the $(J=0,T=1)$ channel %the JW$^\prime$ interaction leads to a much more rigid structure for the even-nuclei (see Fig. %\ref{fig:pairs}). This rigidification is of course not realistic and other TBME and SPE should %have been modified at the same time, but the low-energy structure of the odd-nuclei is obviously %less concerned by this lack due to the presence of the unpaired proton. 
Then the conclusion from the present work is straightforward~: the first excited levels of the proton-deficient $N=50$ odd-isotones down to $Z=31$ can be described in a coherent way by assuming~: (\textit{i}) an inert $^{78}$Ni core (\textit{ii}) rather clean proton $1f_{5/2}^{n}$ configuration states or $2p_{3/2}$ q.p. states for the lowest states (\textit{iii}) a reasonable pairing strength between the $1f_{5/2}$ protons and (\textit{iv}) (from the success of the jj4b interaction) that the diffusion of pairs of protons across $Z=28$, should it exists, remains limited. We implicitly confirm that the $\pi 1f_{5/2}$ single particle state is lower in energy that $\pi 2p_{3/2}$ for very neutron-rich nuclei which corresponds to an inversion as respect to the known order at stability. So far both $Z=28$ and $N=50$ appear to be effective gaps in the region of $^{81}$Ga, only three protons away from $^{78}$Ni.

\end{document}